\documentclass[preprint,12pt]{elsarticle}

\usepackage{amssymb}
\usepackage{amsmath}

\usepackage{amsthm}
\newtheorem{definition}{Definition}

\usepackage{amsmath, amsfonts}
\usepackage{array}
\usepackage{graphicx}
\usepackage{url}
\usepackage{multirow}
\usepackage{makecell}
\usepackage{xcolor}
\usepackage{listings}
\usepackage{tabularx}

\usepackage[hidelinks]{hyperref}
\usepackage{subcaption}
\usepackage{algorithm}
\usepackage{algorithmic}

\lstdefinelanguage{hlpsl}{
  morekeywords={request, role, agent, symmetric_key, channel, played_by, def, local, init, transition, State, text, new, secret, witness, end, goal, protocol_id, composition},
  sensitive=true,
  morecomment=[l]{//},
  morestring=[b]",
}

\journal{Nuclear Physics B}

\begin{document}

\begin{frontmatter}



\title{Quantum-Resistant Authentication Scheme for RFID Systems Using Lattice-Based Cryptography}


\author[inst1]{Vaibhav Kumar}
\ead{vaibhavkumar260702@gmail.com}

\author[inst1]{Kaiwalya Joshi}
\ead{kaiwalyajoshi17@gmail.com}

\author[inst1]{Bhavya Dixit\corref{cor1}}
\ead{22d0549@iitb.ac.in}

\author[inst1]{Gaurav S. Kasbekar}
\ead{gskasbekar@ee.iitb.ac.in}

\cortext[cor1]{Corresponding author}

\affiliation[inst1]{
  organization={Department of Electrical Engineering, Indian Institute of Technology Bombay},
  city={Mumbai},
  postcode={400076},
  state={Maharashtra},
  country={India}
}

\begin{abstract}
We propose a novel quantum-resistant mutual authentication scheme for radio-frequency identification (RFID) systems. Our scheme uses lattice-based cryptography and, in particular,  achieves quantum-resistance  by leveraging the hardness of the inhomogeneous short integer solution (ISIS) problem. In contrast to prior work, which assumes that the reader-server communication channel is secure, our scheme is secure  even when both the reader-server and tag-reader communication channels are insecure. Our proposed protocol provides robust security against man-in-the-middle (MITM), replay, impersonation, and reflection attacks, while also ensuring unforgeability and preserving anonymity. We present a detailed security analysis, including  semi-formal analysis and formal verification using the Automated Validation of Internet Security Protocols and Applications (AVISPA) tool. In addition, we analyze the storage, computation, and communication costs of the proposed protocol and compare its security properties with those of existing protocols, demonstrating that our scheme offers strong security guarantees. To the best of our knowledge, this paper is the first quantum-resistant authentication protocol for RFID systems that comprehensively addresses the insecurity of both the reader-server and tag-reader communication channels.
\end{abstract}




\end{frontmatter}



\section{Introduction}
\label{sec1}


The Internet of Things (IoT) is evolving rapidly, revolutionizing human interactions with the physical world by embedding connectivity into sensors, smart devices, and everyday objects \cite{article}. By linking these components to the Internet, the IoT enables a seamless integration of  digital and physical environments and supports a broad spectrum of applications, including smart healthcare \cite{healthcare,kashani2021systematic}, environmental sensing, precision farming, intelligent transportation \cite{transportation}, and home automation \cite{article}. This widespread integration of IoT into various aspects of daily life results in several security challenges \cite{article1}. Establishing a secure IoT ecosystem requires robust authentication of all participating entities to ensure that only legitimate users and devices engage in network communication \cite{10.1109/MITP.2017.3680960,mahbub2020progressive}.

A key IoT technology is radio-frequency identification (RFID)  systems, which use radio waves to automatically identify and track objects. They consist of tags attached to items and readers that detect the tags within their range, enabling applications such as access control, inventory management, and asset tracking \cite{1593568}. RFID facilitates fast and contactless data transfer, making it widely used across industries. In RFID systems, mutual authentication schemes play a crucial role in ensuring device security \cite{s18103584,electronics12132990}.

RFID systems consist of three components: the RFID tag, the reader, and the server (see Fig. \ref{system_model}). Each tag, which carries a unique identifier (ID), is associated with an object, allowing the reader to identify the object through the tag \cite{6177411}. If the reader needs additional information about the tag, it can retrieve the information from the server. During the authentication process, mutual authentication is performed between the server and the reader, as well as between the server and the tag \cite{s18103584}.

Although many RFID authentication protocols have been proposed \cite{s18103584,electronics12132990,Chien2010,inproceedings,CHO201558,DASS2016100,dong2018cloud,RePEc:sae:intdis:v:15:y:2019:i:7:p:1550147719862223,KUMAR20198,KUMARI2020102443,provable,hwang2023improvement,bahache2025efficient,timouhin2023new,maurya2023mds,gao2014ultralightweight}, the vast majority rely on classical cryptographic assumptions and are therefore vulnerable to adversaries equipped with quantum computers. Quantum computers are expected to dramatically enhance computational power by exploiting quantum phenomena such as superposition and entanglement, enabling solutions to certain problems that are intractable for classical computers \cite{Preskill2018quantumcomputingin}. Shor's algorithm, in particular, enables the efficient solution of difficult problems, such as the computation of discrete logarithms and integer factorization, on quantum computers \cite{doi:10.1137/S0036144598347011}. In addition, Grover's algorithm provides broad asymptotic speed-ups to many kinds of brute-force attacks on symmetric-key cryptography, including collision attacks and pre-image attacks \cite{grover1996fastquantummechanicalalgorithm}.  Public key cryptosystems such as Rivest–Shamir–Adleman (RSA), Digital Signature Algorithm (DSA), and Elliptic Curve Digital Signature Algorithm (ECDSA), which were traditionally considered secure, will become vulnerable in the post-quantum era \cite{nistpqcreport2016}. As a result,  novel authentication schemes, which are quantum-resistant, must be developed for post-quantum RFID systems. Cryptographic methods based on hash functions, codes, lattices, multivariate polynomials, and isogenies are believed to be secure, as no polynomial-time algorithms have been proposed to solve these problems on quantum computers \cite{nistpqcreport2016}.

Conventionally, most authentication schemes are designed under the assumption that the communication channel between the tag and the reader is insecure and prone to attacks, whereas the channel between the reader and the server is secure. However, in increasingly complex IoT environments such as distributed and cloud environments, even the reader-server communication channel can be vulnerable to attacks ~\cite{a2,a3}.

Numerous RFID mutual authentication protocols have been proposed using techniques such as Pseudo Random Number Generator (PRNG), hash functions, and Elliptic Curve Cryptography (ECC) to achieve classical security \cite{Chien2010, inproceedings, CHO201558, DASS2016100, KUMAR20198, KUMARI2020102443,hwang2023improvement,bahache2025efficient,timouhin2023new}. However, many of these schemes either lack scalability and/ or identity privacy, or rely on assumptions such as the availability of a secure channel between the reader and the server; also, critically, most are not quantum-resistant. While several recent works have introduced post-quantum secure schemes using lattice-based  and ring learning with errors (LWE)-based constructions \cite{soni2025lb,fathenojavan2025post,base, ghosh, chen2025security}, they assume the reader-server channel to be secure. Our work improves upon these by providing a lattice-based quantum-resistant mutual authentication protocol that does not require a trusted reader-server channel. Among the various post-quantum cryptographic families, we choose the lattice-based family because it offers a balanced trade-off between security and efficiency, unlike code-based schemes, which suffer from large public key sizes~\cite{cryptoeprint:2021/492}, isogeny-based schemes, which have been recently broken~\cite{castryck2022breaking}, and multivariate schemes, which were significantly weakened after attacks on the National Institute of Standards and Technology (NIST) signature finalist Rainbow \cite{10.1007/978-3-031-15979-4_16}.

The contributions of this paper are as follows. We propose a novel lattice-based mutual authentication scheme for RFID systems, which does not require a trusted reader-server channel and achieves strong security in both classical and post-quantum settings. Our scheme leverages the hardness of the inhomogeneous short integer solution (ISIS) problem \cite{isis,wang2014lattice} to achieve quantum-resistance.  The proposed scheme offers robust security against various threats, including man-in-the-middle (MITM) attacks, replay attacks, impersonation attacks, and reflection attacks. Additionally, it ensures unforgeability and preserves user identity privacy. We present semi-formal security proofs that demonstrate the resilience of the scheme against MITM attacks, replay attacks, impersonation attacks, and reflection attacks, and show that it achieves unforgeability and identity privacy. Furthermore, we conduct formal verification of our proposed scheme using the Automated Validation of Internet Security Protocols and Applications (AVISPA) tool \cite{avispa}. Finally, we analyze the proposed scheme in terms of storage, computation, and communication costs. To the best of our knowledge, this paper is the first to propose a quantum-resistant  authentication scheme for RFID systems that is secure  even when both the reader-server and tag-reader communication channels are insecure. 

The remainder of this paper is organized as follows. Section~\ref{related_works} presents a review of related work. Section~\ref{system model and problem_formulation} outlines the system model, adversary model, and the security goals considered in the design of the proposed protocol. In Section~\ref{proposedscheme}, we describe our proposed scheme, beginning with a brief review of the ISIS problem and then explaining the setup phase and the authentication phase of the proposed scheme. Section~\ref{securityanalysis} provides a semi-formal security analysis of our proposed protocol and Section~\ref{formal_verification} provides a formal verification of the proposed scheme using the AVISPA tool \cite{avispa}. Section~\ref{performance_evaluation} evaluates the performance of our scheme in terms of the storage cost, communication cost, and computational cost. Section \ref{comparison} compares the security properties of our proposed scheme with those of existing schemes. Finally, Section~\ref{conclusion} concludes the paper and outlines some directions for future research.

\section{Related Work} \label{related_works}

Several authentication schemes have been proposed for RFID systems, which can broadly be categorized based on the underlying cryptographic primitives and the security assumptions regarding the reader-server communication channel.

From the perspective of cryptographic techniques, in \cite{Chien2010},  a basic exclusive OR (XOR) and PRNG-based protocol was introduced. The work \cite{inproceedings} improved on this with a cyclic redundancy check (CRC), and  \cite{CHO201558} used hash functions to enhance security. Although these schemes provide mutual authentication and resist attacks such as replay and MITM, they lack identity/ location privacy, scalability, and resistance against quantum attacks. Similarly, \cite{DASS2016100} employed PRNG and hash functions to provide classical security, but the proposed scheme did not ensure scalability or post-quantum protection. In \cite{dong2018cloud}, security for reader-cloud communication was added, but the scheme still lacked quantum-resistance.

To provide stronger classical security guarantees, several works have explored more advanced cryptographic constructions. In \cite{hwang2023improvement}, the authors proposed a lightweight near-field communication (NFC) authentication scheme based on modified hash functions, focusing on efficiency, mutual authentication, and resistance to replay and location attacks, while also addressing synchronization issues. While such hash-based schemes offer high performance due to minimal computational overhead, they lack robust cryptographic security against more sophisticated attacks (e.g., impersonation, traceability) and are not quantum-resistant, as Grover's algorithm can accelerate brute-force attacks on hash functions, requiring significantly larger hash output sizes to maintain post-quantum security. In contrast, our lattice-based approach provides strong quantum resistance and comprehensive security features.

In \cite{RePEc:sae:intdis:v:15:y:2019:i:7:p:1550147719862223},  a matrix-based scheme with scalability and location privacy was proposed. Similarly, \cite{KUMAR20198, KUMARI2020102443} introduced elliptic curve cryptography (ECC)-based authentication protocols and formally analyzed their security. However, the protocols proposed in \cite{RePEc:sae:intdis:v:15:y:2019:i:7:p:1550147719862223,KUMAR20198, KUMARI2020102443} provide strong security against classical adversaries but remain vulnerable to quantum attacks.

The works \cite{bahache2025efficient} and \cite{timouhin2023new} exemplify the continued development of ECC-based RFID authentication protocols, often targeting specific applications such as healthcare. These schemes leverage ECC's efficiency for resource-constrained devices, providing mutual authentication, resistance to common classical attacks (e.g., replay, impersonation), and ensuring properties like confidentiality and integrity, with formal verification using tools such as AVISPA. However, these protocols, like all classical cryptographic solutions, are inherently vulnerable to quantum attacks based on Shor's algorithm. Furthermore, while \cite{bahache2025efficient} acknowledges the ``vulnerable wireless channels between tags and readers", and \cite{timouhin2023new} aims for security between ``RFID cards, card readers, and servers", their system models, typical of many ECC-based schemes, often implicitly or explicitly assume a secure reader-server channel, a critical vulnerability that our quantum-resistant scheme explicitly addresses by securing both communication links. 

The work \cite{maurya2023mds} proposed an ultra light-weight authentication protocol using group homomorphism and maximum distance separable (MDS) codes. Similarly, an ultra light-weight protocol (LPCP) \cite{gao2014ultralightweight} integrating CRC-16 and permutations was proposed in. While these schemes achieve high efficiency suitable for constrained environments, they rely on linear algebraic constructions without cryptographic hardness guarantees, making it vulnerable to advanced adversaries and unsuitable for post-quantum settings.

To address the limitations of classical cryptography, several post-quantum RFID authentication schemes \cite{kumar2022comprehensive} have been proposed. In \cite{soni2025lb}, the authors  introduced a Bi-ISIS-based lattice protocol, LB-RFID, for resource-constrained RFID systems. It offers provable security (validated using the real-or-random (ROR) and  Canetti-Krawczyk (CK) models and Scyther), low communication and storage overhead, and strong security features such as mutual authentication, anonymity, forward secrecy, and resistance to various attacks. Similarly, \cite{fathenojavan2025post} presented LBAA, an ISIS-based lightweight protocol for IoT RFID, prioritizing enhanced anonymity and untraceability. It achieves unforgeability, scalability, mutual authentication, availability, and perfect forward secrecy, and  formal verification was done. The schemes proposed in \cite{soni2025lb} and \cite{fathenojavan2025post} are quantum-resistant. In addition, \cite{base} proposed a lattice-based RFID authentication protocol with quantum resistance, while \cite{ghosh} presented a Ring-LWE based quantum-secure authentication scheme. These approaches demonstrate the potential of lattice-based cryptography for building RFID systems that remain secure in the post-quantum era.

Another important consideration in RFID authentication protocol design is the assumption regarding the security of the reader-server communication channel. Many existing schemes implicitly or explicitly assume that this channel is secure. In \cite{provable}, a pseudo-random generator and vector dot product were used to resist common attacks, but it was assumed that the reader-server channel is secure. In addition, the proposed scheme is not post-quantum secure. Also, \cite{soni2025lb}, \cite{fathenojavan2025post}, \cite{maurya2023mds}, \cite{base}, \cite{ghosh} assume that the reader and server communicate over a secure channel. In contrast, our proposed scheme does not require a trusted reader-server channel and provides strong security in both classical and post-quantum settings.

\section{System Model, Problem Formulation, Adversary Model, and Security Goals}\label{system model and problem_formulation}

\subsection{System Model}

\begin{figure}[t]
    \centering
    \includegraphics[width=1\linewidth]{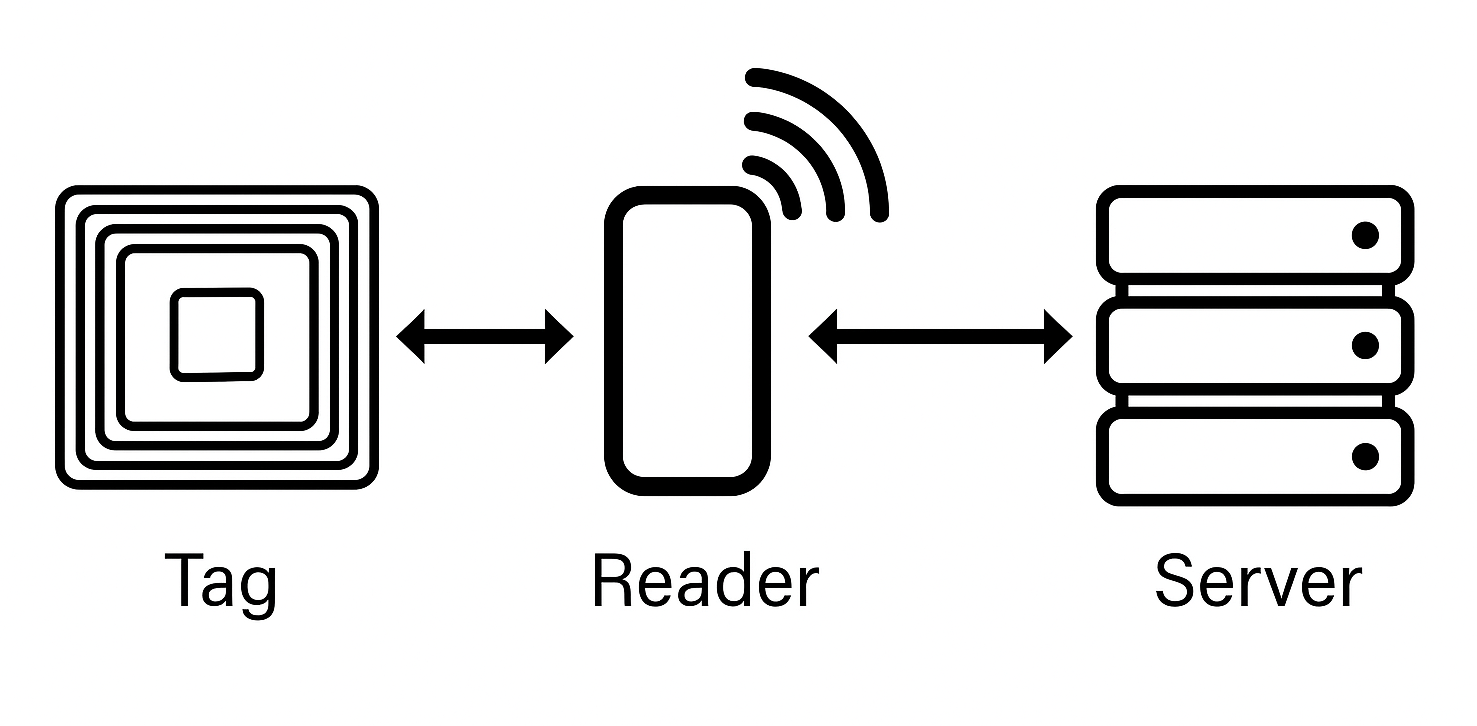}
    \caption{The figure shows our system model.}
    \label{system_model}
\end{figure}

Our system model of an RFID system consists of three primary entities: \textit{Tag}, \textit{Reader}, and \textit{Server} (see  Fig. \ref{system_model}). All communication in the system flows through the Reader, which acts as an intermediary between the Tag and the Server. Communication channels may be either wired or wireless and are not assumed to be secure. Each Tag and Reader is provisioned with a unique identity and secret parameters by the Server during an initial setup or registration phase. The Server, deployed in a secure and trusted environment, maintains a database of registered devices and is assumed to be fully secure and impervious to adversarial compromise or impersonation.

\subsection{Adversary Model}

We adopt the Dolev-Yao (DY) adversary model~\cite{DY}, which is widely used in the analysis of cryptographic protocols \cite{9720852,1443199,1310742}. Recall that an adaptive  DY adversary can intercept, modify, and impersonate entities based on what it observes in the current session. In contrast, a static DY adversary does not learn in real-time and cannot change its strategy on the fly using intercepted messages. While a static adversary is weaker, an adaptive adversary is more flexible and smarter, making it a stronger threat model. In this paper, we allow the DY adversary to be adaptive. The protocol participants are assumed to behave honestly, while the communication channels are considered insecure. The adversary, say $\mathcal{A}$, is capable of performing various operations, including the following: 
\begin{enumerate}[i.]
    \item $\mathcal{A}$ can obtain any message from any of the public channels in the system.
    \item $\mathcal{A}$ is a legitimate user of the system who can receive and send messages over any public channel.
    \item $\mathcal{A}$ can eavesdrop, intercept, and modify messages between other entities.
    \item $\mathcal{A}$ can impersonate other entities to receive and send messages.
\end{enumerate}

Note that both the communication channels- between the tag and the reader, and between the reader and the server- are assumed to be insecure and susceptible to eavesdropping, message tampering, and replay attacks. 

\subsection{Security Goals}\label{security_goals} Our goal is to design an authentication protocol that achieves several critical security objectives. First, it must ensure mutual authentication between the tag, the reader, and the server, allowing all the parties to verify each other's identities. The proposed scheme must ensure secure communication over the tag-reader and reader-server channels, both of which are considered insecure. It should be robust against various attacks, including impersonation \cite{Mitrokotsa2010}, MITM \cite{sivasankari2022detection}, replay \cite{Mitrokotsa2010}, and reflection attacks \cite{lee2021countermeasures}. The scheme must also preserve identity privacy \cite{anonymity}, ensure unforgeability \cite{cryptoeprint:2017/1221}, unlinkability \cite{8620304}, and support scalability \cite{8813679}. Furthermore, it should be resilient to quantum attacks.

We make the following assumptions:
\begin{enumerate}
   
    \item \textbf{Secure Server:} The backend server is fully secure-- it has cryptographic integrity, i.e., the server's internal computations are performed honestly and cannot be tampered with by an outside party, and physical safety, i.e., an attacker cannot physically break into the server. In particular, the sensitive parameters stored within the server’s database cannot be compromised through physical access, insider threats, or external breaches of the server itself. This assumption is common in the literature, as most existing RFID authentication protocols rely on a secure and trusted backend database \cite{Chien2010, inproceedings, CHO201558, DASS2016100}.

    \item \textbf{Hardness of ISIS Problem:} The ISIS problem is hard and no efficient classical or quantum algorithm exists to solve it  \cite{isis,wang2014lattice}.

\end{enumerate}

\section{Proposed Authentication Scheme} \label{proposedscheme}

\subsection{Overview}

The proposed authentication scheme is designed to provide authentication between the tag, the reader, and the server. The protocol operates in two phases: the Setup Phase (Section \ref{SSC:setup:phase}) and the Authentication Phase (Section \ref{SSC:authentication:phase}). During the Setup Phase, secret parameters are initialized and securely distributed. In the Authentication Phase, the tag, the reader, and the server exchange messages and perform verification and related operations to authenticate each other. We prove the correctness of different steps of our proposed protocol in Section \ref{SSC:correctness:protocol:steps}.  Our proposed scheme is based on the quantum-resistant ISIS problem, which is reviewed in Section \ref{isis}.

\subsection{ISIS Problem}\label{isis}

Let \(\mathbb{Z}\) denote the set of all integers. For a positive integer $q$, let \(\mathbb{Z}_q = \{0, 1, \ldots, q-1\}\). Also, recall that for \(p \geq 1\), the \(\ell_p\) norm of a vector \(x = (x_1, \ldots, x_m)\), where $m$ is a positive integer, is defined as \(\|x\|_p = \left( \sum_{i=1}^m |x_i|^p \right)^{1/p}\) \cite{rudin1987real}.

\begin{definition}[ISIS Problem~\cite{isis}]
Let $m$ and $n$ be positive integers. Given a modulus \( q \), a vector \( y \in \mathbb{Z}_q^n \),  and a matrix \( A \in \mathbb{Z}_q^{n \times m} \), where the entries of $y$ and $A$ are  independently selected at random using a  uniform distribution over \(\mathbb{Z}_q \), the \( \text{ISIS}^{p}_{n,m,q,\sigma} \) problem in the \( \ell_p \) norm is to find a vector \( x \in \mathbb{Z}^m \) such that \( \|x\|_p \leq \sigma \) and
\[
Ax = y \mbox{ mod } q.
\]
Here, \(\sigma\) is a  bound on the \( \ell_p \) norm that restricts the solution vector \( x \) to have only small integer entries.    
\end{definition}

The hardness of the \( \text{ISIS}^{p}_{n,m,q,\sigma} \) problem has been established through worst-case to average-case reductions from standard lattice problems, such as the Shortest Independent Vector Problem (SIVP) \cite{isis}.

We adopt the parameter set proposed in \cite{Cayrel20123363}. In particular, we choose \( m = \Theta(n \log n) \) and a prime modulus \( q = \mathcal{O}(n^{2.5} \log n) \). This specific relationship between $n$, $m$, and $q$ ensures that solving the ISIS problem is on average at least as difficult as solving the $\mbox{GapSVP}_\gamma$ problem in the worst case \cite{Cayrel20123363}.



\subsection{Setup Phase}
\label{SSC:setup:phase}
Table \ref{table:notations} provides a summary of the notation that we use. All computations in the proposed scheme are performed modulo \( q \). \( S_m \)
denotes the set of all permutations on \( m \) elements and any permutations \( \Sigma_r, \Sigma_t \in S_m \) are linear operations. \( P_t \) and  \( P_r \) are \( m \times m \) binary matrices. Also, \( A_t \) and  \( A_r \) are
matrices chosen from \( \mathbb{Z}_q^{n \times m} \).

In our proposed scheme, \( A_t \) and \( y_t \) are shared keys between the tag
and the server. Similarly, \( A_r \) and \( y_r \) are shared keys between the reader and the server. Two hash functions are utilized; each RFID tag and reader is embedded with these hash functions, which are also known to the server. The first hash function, denoted by \( H_1 \), is used for standard hashing operations. The second hash function, denoted by \( H_2 \), generates a hash of length \( m - n \), where \( m \) and \( n \) represent the number of columns and rows, respectively, in the matrix \( A_t \). The function \( H_2 \) is specifically employed for concatenation and for adjusting the length of vectors within the scheme.

 First, the server generates \( P_t \), \( A_t \), and \( y_t \) for each tag and \( P_r \), \( A_r \), and \( y_r \) for each reader. Each tag (respectively, reader) has a unique ID denoted by \( x_t \) (respectively,  \( x_r \)).   
 These IDs are non-zero solution vectors of the ISIS problem and they satisfy the following conditions:
\begin{align*}
\|x_t\|_p &\leq \sigma, \\
\|x_r\|_p &\leq \sigma, \\
        A_r x_r &= y_r \, (\mbox{mod } q), \\
        A_t x_t &= y_t \, (\mbox{mod } q).
\end{align*}
The values of \( x_t \), \( P^{-1}_t \), and \( y_t \) (respectively, \( x_r \), \( P^{-1}_r \), and \( y_r \)) for each tag (respectively, reader) are stored in the server’s database.  The server also computes the results of the multiplications \( A_tP^{-1}_t \) and \( A_rP^{-1}_r \) to verify the commitment values.


\begin{figure*}
    \centering
    \fbox{\includegraphics[width=1\linewidth]{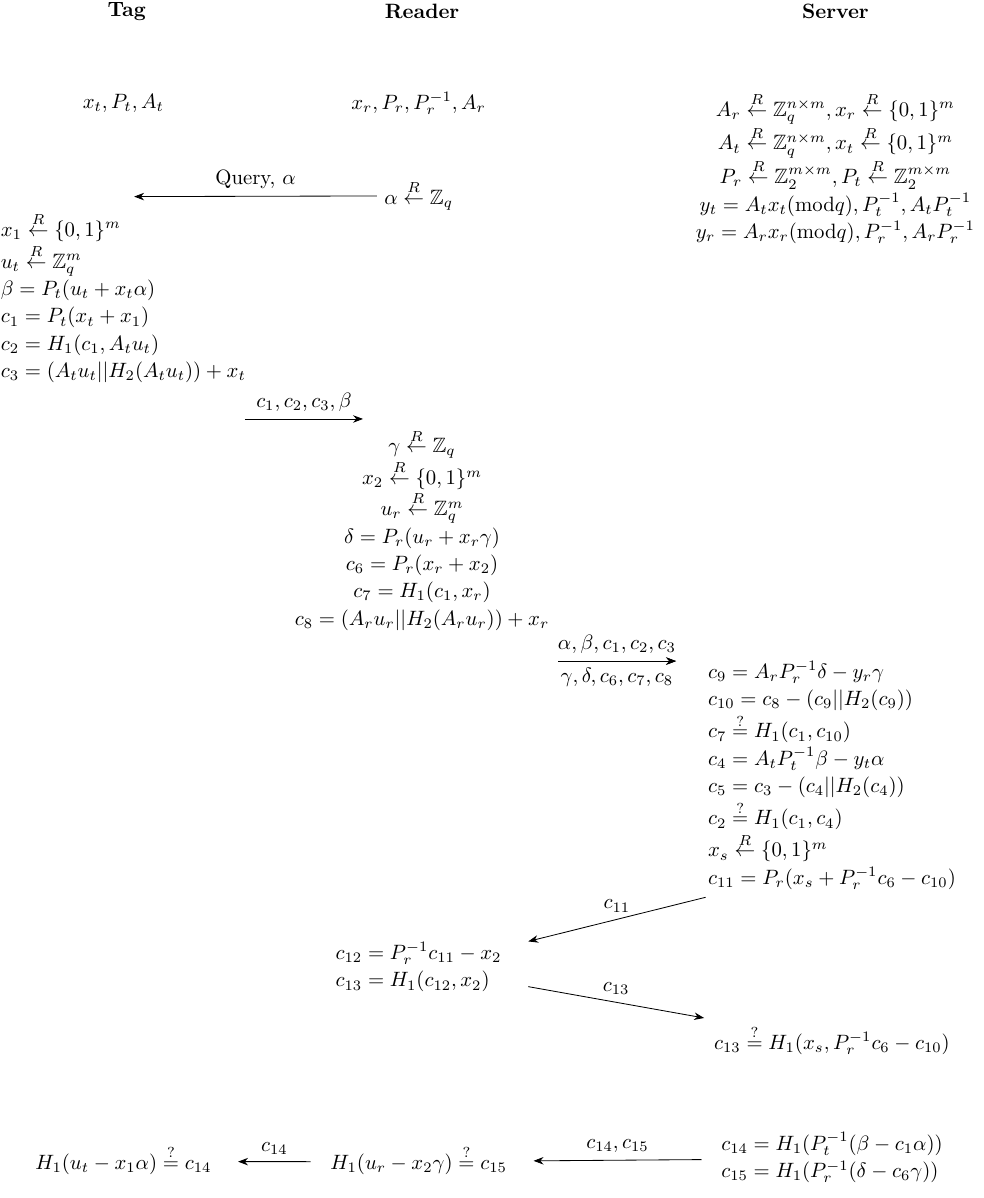}}
    \caption{The figure shows the proposed authentication scheme. The notation $x \xleftarrow{R} S$ denotes that $x$ is chosen uniformly at random from the set $S$. The notation $x \stackrel{?}{=} y$ indicates that the recipient checks whether the two values $x$ and $y$ are equal and aborts the protocol if the test fails.}
    \label{fig:scheme}
\end{figure*}

\begin{table}[t]
\centering
\caption{The table provides a summary of the notation we use in this paper.}
\label{table:notations}
\setlength{\tabcolsep}{6pt}
\renewcommand{\arraystretch}{1.1}
\begin{tabular}{ll}
\hline
\textbf{Notation} & \textbf{Definition} \\ 
\hline
\( x_t \) & Tag ID \\ 
\( x_r \) & Reader ID \\ 
\( A_t, P_t \) & Tag’s secret keys \\ 
\( A_r, P_r \) & Reader’s secret keys \\ 
\( x_1, u_t \) & Random vectors from tag \\ 
\( x_2, u_r \) & Random vectors from reader \\ 
\( x_s \) & Random vector from server \\ 
\( S_m \) & Set of all permutations  \\ 
& on \( m \) elements \\ 
\( y_t, y_r \) & Keys derived by server \\ 
\( H_1, H_2 \) & Hash functions \\ 
\( \alpha, \gamma \) & Reader-generated random values \\ 
\( n, m, q, p \) & ISIS security parameters \\ 
\( c_1, c_2, c_3, \beta \) & Values computed by tag \\ 
\( c_6, c_7, c_8, \delta, c_{12}, c_{13} \) & Values computed by reader \\ 
\( c_4, c_5, c_9, c_{10}, c_{11}, c_{14}, c_{15} \) & Values computed by server \\ 
\hline
\end{tabular}

\end{table}

\subsection{Authentication Phase}
\label{SSC:authentication:phase}

After the values of \(x_t\), \(P_t\), \(A_t\), \(x_r\), \(P_r\), and \(A_r\) have been chosen, the authentication
phase starts. This phase is performed interactively among the tag, reader, and server.
At a high level, this three-party interaction occurs as follows:
\begin{itemize}
    \item First, mutual binding is performed between the reader and the tag, in which the tag and reader generate random vectors ($x_1,u_t$) and ($x_2,u_r$), respectively. Also, commitment values $(c_1,c_2,c_3,c_6,c_7,c_8)$ are generated, which are cryptographically bound to the current session via the random nonces $\alpha$ and $\gamma$.
    \item Second, instead of the tag and reader sending their IDs $x_t$ and $x_r$  directly, they send commitments that allow the server to mathematically extract and verify the IDs $x_r$ and $x_t$ using an ISIS-based derivation of the shared keys $y_r$ and $y_t$ of the reader and tag, respectively.
    \item Finally, a three-way handshake ($c_{12},c_{13},c_{14},c_{15}$) ensures that the server's successful verification of the reader and tag is securely informed back to the reader and tag, which completes the mutual authentication process.
\end{itemize}

More elaborately, the following steps are followed in the authentication phase (see Fig. \ref{fig:scheme}).  

\hypertarget{step1}{}\textbf{Step 1:} The reader chooses a random value \(\alpha \in \mathbb{Z}_q\) and sends it to
the tag along with a query signal.

\hypertarget{step2}{}\textbf{Step 2:} In this step, we have two sub-steps:
\begin{itemize}
\hypertarget{step2a}{}\item \textbf{Step 2.1:} The tag randomly chooses the secret key \( u_t \in \mathbb{Z}_q^m \). In addition, the tag randomly selects a value \( x_1 \in \{0, 1\}^m \).
    \hypertarget{step2b}{}\item \textbf{Step 2.2:} The tag receives the random value \(\alpha\) and computes four
    values:
    \begin{align*}
\beta &= P_t (u_t + x_t \alpha), \\
c_1 &= P_t (x_t + x_1), \\
c_2 &= H_1(c_1, A_t u_t), \\
c_3 &= (A_tu_t \,\|\, H_2(A_t u_t)) +  x_t,
\end{align*}
where  \(\,\|\) denotes the concatenation operation. The tag sends these four commitment values to the reader.

\end{itemize}

\hypertarget{step3}{}\textbf{Step 3:} In this step, we have two sub-steps:
\begin{itemize}
\hypertarget{step3a}{}\item \textbf{Step 3.1:} The reader chooses a random value \(\gamma \in \mathbb{Z}_q\). The reader also randomly chooses the secret key \( u_r \in \mathbb{Z}_q^m \). In addition, the reader randomly selects a value \( x_2 \in \{0, 1\}^m \).

    \hypertarget{step3b}{}\item \textbf{Step 3.2:} The reader computes four
    values:
    \begin{align*}
\delta &= P_r (u_r + x_r \gamma), \\
c_6 &= P_r (x_r + x_2), \\
c_7 &= H_1(c_1, x_r), \\
c_8 &= (A_ru_r \,\|\, H_2(A_r u_r)) + x_r.
\end{align*}

    The reader sends the ten commitment values \(\alpha\), \(\beta\), \(c_1\), \(c_2\), \(c_3\), \(\gamma\), \(\delta\), \(c_6\), \(c_7\), and \(c_8\) to the server.

\end{itemize}

\hypertarget{step4}{}\textbf{Step 4:} In this step, we have three sub-steps:
\begin{itemize}
\hypertarget{step4a}{}\item \textbf{Step 4.1:} The server computes

\[
c_9 = A_r P_r^{-1} \delta - y_r \gamma \quad \text{and} \quad c_{10} = c_8 - (c_9 \,\|\, H_2(c_9))
\]
to obtain the ID \(x_r\). Then, the server checks whether
\[
c_7 = H_1(c_1, c_{10}).
\]
If this condition is successfully verified, then the server infers that it is communicating with a valid reader. Otherwise, the authentication process is terminated.

\hypertarget{step4b}{}\item \textbf{Step 4.2:} In this step, the server computes
\[
c_4 = A_t P_t^{-1} \beta - y_t \alpha \quad \text{and} \quad c_5 = c_3 - (c_4 \,\|\, H_2(c_4))
\]
to obtain the ID \(x_t\). Then, the server checks whether
\[
c_2 = H_1(c_1, c_4).
\]
If this condition is successfully verified, then the server concludes that it is communicating with a valid tag. Otherwise, the authentication process is terminated.

\hypertarget{step4c}{}\item \textbf{Step 4.3:} The server randomly selects a value \( x_s \in \{0, 1\}^m \), calculates 
\[
c_{11} = P_r(x_s + P_r^{-1}c_6-c_{10}),
\]
and sends \(c_{11}\) to the reader. 
\end{itemize}

\hypertarget{step5}{}\textbf{Step 5:} The reader receives \(c_{11}\),  computes
\begin{align*}
    c_{12} = P_r^{-1}c_{11}-x_2 \text{  and  }
    c_{13} = H_1(c_{12},x_2),
\end{align*}
and sends \(c_{13}\) to the server.

\hypertarget{step6}{}\textbf{Step 6:} This step has two sub-steps.

\begin{itemize}
    \item \hypertarget{step6a}{}\textbf{Step 6.1:} The server authenticates the reader and tag by checking whether
\[
c_{13}  = H_1(x_s, P_r^{-1}c_6-c_{10})
\]
holds.

\item \hypertarget{step6b}{}\textbf{Step 6.2:}  The server generates
\[
c_{14} = H_1(P_t^{-1} (\beta - c_1 \alpha)) \quad
\text{and} \quad
c_{15} = H_1(P_r^{-1} (\delta - c_6 \gamma))
\]
and sends them to the reader.
\end{itemize}

\hypertarget{step7}{}\textbf{Step 7:} The reader receives \(c_{14}\) and \(c_{15}\) and checks whether
\[
H_1(u_r - x_2 \gamma) = c_{15}.
\]
If this equation does not hold, then the authentication of the reader by the server fails. The reader also sends \(c_{14}\) to the tag.

\hypertarget{step8}{}\textbf{Step 8:} The tag receives \(c_{14}\) and checks whether
\[
H_1(u_t - x_1 \alpha) = c_{14}.
\]
If this equation does not hold, then the authentication of the tag by the reader fails.

\subsection{Correctness of Protocol Steps}
\label{SSC:correctness:protocol:steps}
In this section, we prove the correctness of different steps of our proposed protocol. 

\noindent \textbf{Correctness I:} 
The server computes \( c_9 \) in \hyperlink{step4a}{Step 4.1} using \( \delta \), which is sent by the reader. This value can be shown to equal $A_r u_r$ as follows:
\begin{align*}
        c_9 &= A_r P^{-1}_r \delta - y_r \gamma \\
        &= A_r P^{-1}_r (P_r u_r + P_r x_r \gamma) - y_r \gamma \\
        &= A_r u_r + A_r x_r \gamma - y_r \gamma \\
        &= A_r u_r + A_r x_r \gamma - A_r x_r \gamma \tag*{(since \( y_r = A_r x_r \))} \\
        &= A_r u_r.
\end{align*}

\noindent \textbf{Correctness II:}
Using the following equations, the server obtains the ID of the reader in \hyperlink{step4a}{Step 4.1}:
\begin{align*}
        c_{10} &= c_8 - (c_9 \,\|\, H_2(c_9)) \\
        &= (A_r u_r \,\|\, H_2(A_ru_r)) + x_r - (A_r u_r \,\|\, H_2(A_ru_r)) \\
        &= x_r.
\end{align*}
Note that the server generates \( c_9 \) in the previous step, whereas \( c_8 \) is sent by the reader in \hyperlink{step3b}{Step 3.2}.

\noindent \textbf{Correctness III:} 
The server computes \( c_4 \) in \hyperlink{step4b}{Step 4.2} using \( \beta \), which is sent by the tag. This value can be shown to equal $A_t u_t$ as follows: 
\begin{align*}
        c_4 &= A_t P^{-1}_t \beta - y_t \alpha \\
        &= A_t P^{-1}_t (P_t u_t + P_t x_t \alpha) - y_t \alpha \\
        &= A_t u_t + A_t x_t \alpha - y_t \alpha \\
        &=  A_t u_t + A_t x_t \alpha - A_t x_t \alpha 
        \tag*{(since \( y_t = A_t x_t \))} \\
        &= A_t u_t.
\end{align*}

\noindent \textbf{Correctness IV:}
Using the following equations, the server obtains the ID of the tag in \hyperlink{step4b}{Step 4.2}: 
\begin{align*}
        c_5 & = c_3 - (c_4 \,\|\, H_2(c_4)) \\
        &= (A_t u_t \,\|\, H_2(A_tu_t)) + x_t - (A_t u_t \,\|\, H_2(A_tu_t)) \\
        &= x_t. 
\end{align*}
Note that the server generates \( c_4 \) in the previous step, whereas \( c_3 \) is sent by the tag in \hyperlink{step2b}{Step 2.2}.

\noindent \textbf{Correctness V:}
The server computes $c_{11}$ in \hyperlink{step4c}{Step 4.3} using $c_6$ sent by the reader in \hyperlink{step3b}{Step 3.2} and $x_s$. This value can be shown to equal  $P_r(x_s + x_2)$ as follows: 
\begin{align*}
        c_{11} &= P_r(x_s + P_r^{-1}c_6-c_{10}) \\
        &= P_r(x_s + P_r^{-1}P_r(x_r + x_2)-x_r) \tag*{(since \(  c_{10} = x_r \)  (see Correctness II))} \\
        &= P_r(x_s + x_2).
\end{align*}

\noindent \textbf{Correctness VI:}
The reader computes $c_{12}$ in \hyperlink{step5}{Step 5} using $c_{11}$, which is sent by the server in \hyperlink{step4c}{Step 4.3}. The reader thus obtains $x_s$. This can be shown as follows:
\begin{align*}
        c_{12} &= P_r^{-1}c_{11}-x_2 \\
        &= P_r^{-1}P_r(x_s + x_2)-x_2 \\
        &= x_s.
\end{align*}

\noindent \textbf{Correctness VII:}
The server computes $c_{14}$ in \hyperlink{step6b}{Step 6.2} using $c_1,\beta$, and $\alpha$, which were sent by the tag. This value can be shown to equal  $H_1(u_t - x_1\alpha)$ as follows:
\begin{align*}
        c_{14} &= H_1(P_t^{-1}(\beta - c_1\alpha)) \\
        &= H_1(P_t^{-1}(P_t(u_t + x_t\alpha) - P_t(x_t + x_1)\alpha) \\
        &= H_1(P_t^{-1}P_t(u_t + x_t\alpha - x_t\alpha - x_1\alpha)) \\
        &= H_1(u_t - x_1\alpha).
\end{align*}

\noindent \textbf{Correctness VIII:}
The server computes $c_{15}$ in \hyperlink{step6b}{Step 6.2} using $c_6,\delta$, and $\gamma$, which were sent by the reader. This value can be shown to equal $H_1(u_r - x_2\gamma)$ as follows:
\begin{align*}
        c_{15} &= H_1(P_r^{-1}(\delta - c_6\gamma)) \\
        &= H_1(P_r^{-1}(P_r(u_r + x_r\gamma) - P_r(x_r + x_2)\gamma) \\
        &= H_1(P_r^{-1}P_r(u_r + x_r\gamma - x_r\gamma - x_2\gamma)) \\
        &= H_1(u_r - x_2\gamma).
\end{align*}

\section{Security Analysis} \label{securityanalysis}

In this section, we present a semi-formal security analysis of the proposed scheme. In particular, we demonstrate the scheme's resilience against a range of well-known cryptographic attacks.

\subsection{MITM Attacks}\label{mitm}
In the MITM attack, an attacker secretly intercepts and relays messages, possibly after modification, between two parties who believe they are communicating directly with each other \cite{sivasankari2022detection}. We now show that the proposed scheme is resistant to MITM attacks. To account for the worst case, we assume that both communication channels (reader-server and tag-reader) are insecure. 

Suppose an attacker eavesdrops on either of the two communication channels and attempts to intercept or modify the transmitted values. Such an attack will fail because the commitment messages in the proposed scheme are generated using session-specific random values such as $u_r$, $x_2$, $x_1$, and nonces $\alpha$, $\beta$, $\gamma$, and $\delta$. These values ensure that each authentication session is cryptographically independent, preventing replay or manipulation of previously observed messages. Moreover, both the reader and the tag possess unique identities $(x_r, x_t)$, which are implicitly verified during the authentication process. Consequently, any attempt by the attacker to alter transmitted values or inject forged messages will cause the authentication checks to fail. More specifically, the following scenarios illustrate the robustness of the proposed scheme against MITM attacks when an attacker eavesdrops on the two communication channels.

\begin{itemize}
    \item Reader-Server Channel:
    \begin{enumerate}
        \item If an attacker modifies $\beta$, the server's computed $c_4$ will not equal the original $A_t u_t$. Consequently, the check $c_2 \stackrel{?}{=} H_1(c_1, c_4)$ will fail.
        \item If the adversary modifies $\delta$ or $\gamma$, the server's recovery of $x_r$ via $c_{10}$ fails. This leads to a failure of the server's verification equation: $c_7 \stackrel{?}{=} H_1(c_1, c_{10})$.
        \item If $c_6$ is modified, the server's final reader authentication check in Step 6.1, $c_{13} \stackrel{?}{=} H_1(x_s, P_r^{-1}c_6 - c_{10})$, will fail.
    \end{enumerate}
    \item Tag-Reader Channel:
   \begin{enumerate}
       \item If an attacker modifies $c_{14}$ or $\alpha$ as it is passed from the reader to the tag, the tag’s verification $H_1(u_t - x_1 \alpha) \stackrel{?}{=} c_{14}$ will fail.
       \item If $c_{15}$ is modified, the reader’s verification $H_1(u_r - x_2 \gamma) \stackrel{?}{=} c_{15}$ fails, terminating the session immediately.
   \end{enumerate}
\end{itemize}
Therefore, any interception or modification of messages on either the reader-server or tag-reader channel results in failed verification checks, ensuring that the proposed scheme is secure against MITM attacks.

\subsection{Unforgeability} \label{unforge}
Unforgeability ensures that an adversary cannot generate a valid authentication message or response without possessing the legitimate secret credentials \cite{cryptoeprint:2017/1221}. For the unforgeability property to be satisfied, an attacker should not be able to forge the tag's or reader's ID or change any value used in the scheme. Recall that the tag and the reader generate the following values to authenticate: 
\begin{align*}
c_1 &= P_t(x_t + x_1),            & c_6 &= P_r(x_r + x_2), \\
c_2 &= H_1(c_1, A_t u_t),         & c_7 &= H_1(c_1, x_r), \\
c_3 &= (A_t u_t \,\|\, H_2(A_t u_t)) + x_t, & c_8 &= (A_r u_r \,\|\, H_2(A_r u_r)) + x_r, \\
\beta &= P_t(u_t + x_t \alpha),  & \delta &= P_r(u_r + x_r \gamma).
\end{align*}
These commitment values are computed using the tag’s and reader's IDs $x_t$ and $x_r$, respectively, the values $u_t \leftarrow Z_q^m$, $x_1 \leftarrow \{0,1\}^m$, $P_t$, $A_t$, $u_r \leftarrow Z_q^m$, $x_2 \leftarrow \{0,1\}^m$, $P_r$, and $A_r$. Since hash functions, the random values $x_1$ and $x_2$, and the secret keys $u_t$ and $u_r$ are used in the scheme, the attacker cannot derive the correct values of the authentication messages and responses and is not able to forge any legitimate node's messages. Even if these values are leaked, they cannot be reused in another session as new values are computed in each session. Moreover, the use of $c_1$ in the computation of $c_7$ ensures a session-specific connection between tag and reader. Furthermore, due to the hardness of the ISIS problem, on which the proposed scheme is based,  the probability of an attacker breaking it is negligible.

%

\subsection{Replay Attack} \label{replay}

 In a replay attack, an adversary records valid protocol messages and later reuses them in an attempt to deceive the system or gain unauthorized access \cite{Mitrokotsa2010}. Under our proposed scheme, a replay attack is not successful due to the fact that $\alpha,\gamma,x_1, x_2$, and $x_s$ are randomly generated in each session. An eavesdropper can collect all messages exchanged among the tag, the reader, and the server in different sessions. However, if they replay the collected messages, the replay attack will fail as we now explain for different scenarios.
\begin{itemize}
    \item Assume that an eavesdropper sends the commitment values $c_1$, $c_2$, $c_3$, and $\beta$, which were collected during a previous execution of the proposed scheme, to the reader while pretending to be the tag. These values are transmitted to the server by adding the randomly generated values $\alpha$ and $x_1$. Since the reader and the tag randomly generate $\alpha$ and $x_1$ respectively, authentication will fail during the verification of the condition \(H_1(u_t - x_1\alpha) \stackrel{?}{=} c_{14} \).
    \item In the second scenario, assume that an eavesdropper sends the commitment values $c_6$, $c_7$, $c_8$, and $\delta$, which were collected from a previous execution, to the server pretending to be the reader. Authentication will fail while verifying the condition \(H_1(u_r - x_2\gamma) \stackrel{?}{=} c_{15} \) since $\gamma$ and $x_2$ are randomly generated.
\end{itemize}
Thus, the proposed scheme is resistant to replay attacks. 

\subsection{Impersonation Attack} \label{impersonation}
In an impersonation attack, an adversary pretends to be a legitimate entity, such as the tag, reader, or server, in order to gain unauthorized access or disrupt communication \cite{Mitrokotsa2010}. An impersonation attack cannot be made  under our proposed scheme since it is based on the hardness of the ISIS problem. We now show the infeasibility of the impersonation attack in three scenarios: the impersonation of the tag, the reader, and the server.
\begin{itemize}
    \item To impersonate the tag, an attacker needs to obtain the secret keys of the tag, viz., $x_t,P_t$, and $A_t$, from the transmitted messages. However, the transmitted messages $\beta,c_1$, $c_2$, and $c_3$ always use these secret keys along with random numbers $x_1$, $u_t$, or in the inputs to a one-way hash function, which makes it computationally infeasible to obtain these secret keys. Therefore, the attacker must solve the ISIS problem to impersonate the tag, which is computationally infeasible. 
    \item Due to  similar reasons,  reader impersonation is also infeasible. 
    \item To impersonate the server, an attacker must eavesdrop on the transmitted messages and obtain the commitment values. The attacker would then need to respond to the tag's commitment as if they were the server. However, only the server knows the tag's and the reader's secret values $A_t$, $P_t$, $y_t$, $x_t$, $A_r$, $P_r$, $y_r$, and $x_r$. Hence, the attacker would need to solve the ISIS problem to  successfully impersonate the server. Therefore, the proposed scheme is resistant to server impersonation attacks.
\end{itemize}



\subsection{Quantum-Resistance}\label{quantum_resistance}

The proposed scheme employs lattice-based ISIS encryption for secure authentication. It is well established that cryptographic constructions based on the ISIS problem offer strong security guarantees against both classical and quantum adversaries \cite{10.1145/237814.237838,doi:10.1137/S0097539705447360,gentry2008trapdoors}. Hence, our proposed scheme is quantum-resistant.

\subsection{Reflection Attack}\label{reflection}

The reflection attack is a method to attack a challenge–response authentication scheme that uses the same protocol in both directions \cite{lee2021countermeasures}. In our proposed protocol, there is tag-server and reader-server mutual authentication. We now show the infeasibility of the reflection attack for each of these two cases.  
\begin{itemize}
    \item In the tag-server authentication process, the tag transmits the values $c_1, c_2, c_3$, and \(\beta \) to the server. The server, in turn, responds with \( c_{14} \). The structure of these messages is entirely different, making it infeasible for an adversary to generate a valid response without knowing the tag's secret key. Additionally, the use of fresh random vectors in each session prevents the reuse of messages, effectively mitigating the risk of reflection attacks.
    \item During the reader-server authentication process, the reader sends the tuple \( c_6, c_7, c_8, \delta, \gamma \) to the server. The server provides a response \( c_{15} \), which differs structurally and functionally from the reader’s request. This ensures that an attacker cannot successfully impersonate a legitimate reader without possessing the secret key. Furthermore, the use of unique random vectors for each session reinforces the resistance against reflection attacks.
\end{itemize}



\subsection{Anonymity} \label{location}
Anonymity is the process of not revealing a device’s identity to eavesdroppers during communication. It requires leaving no trace marks using which an attacker can trace the communication back to the original devices \cite{anonymity}. No one should be able to obtain information about the tag's or reader's identity or location and an adversary should not be able to map transmitted parameters to unique tags or readers in the system. The proposed scheme satisfies the identity privacy property as the transmitted messages do not leak secret parameters such as $x_t$ and $x_r$. In the proposed scheme, the tag generates the values $\beta$, $c_1$, $c_2$, and $c_3$ to send a response to the reader and the reader generates $\delta$, $c_6$, $c_7$, and $c_8$ to send a response to the server. If the tag or reader were to always send the same values in each session, it would be easy to trace them. Under the proposed scheme, all messages have randomness incorporated in them through random nonces, so an adversary cannot map transmitted messages to a particular tag or reader. 

\subsection{Unlinkability} \label{unlinkability}
Unlinkability ensures that messages from different sessions cannot be mapped to a particular device \cite{8620304}. In the proposed scheme, each session uses fresh random nonces, and all transmitted values incorporate this randomness, making the messages unique across sessions. As a result, an adversary cannot link two different messages to the same entity, thus preserving unlinkability.


\subsection{Scalability} \label{scalable}
Scalability is an essential requirement for RFID-based IoT authentication protocols, especially in environments with a large number of devices and constrained resources~\cite{8813679}. In the proposed scheme, the server, the tag, and the reader each store certain secret parameters during the initial setup phase. These parameters are then reused during authentication, ensuring that the per-authentication computational cost remains constant, and independent of the total number of tags or readers in the system. Therefore, the proposed scheme scales well in practice.

\section{Formal Verification}\label{formal_verification}

To validate the security of the proposed authentication scheme, we performed a formal verification using the AVISPA tool \cite{avispa}. The protocol was specified in High-Level Protocol Specification Language (HLPSL) \cite{hlpsl} and analyzed using the AVISPA ATtack SEarcher (ATSE)  backend \cite{atse}.

Our protocol involves several complex mathematical operations such as modular arithmetic, matrix multiplication, and matrix inversion, which are not natively supported by the AVISPA tool, whose symbolic model primarily handles operations such as XOR and exponentiation. To enable verification within AVISPA, we introduced a set of protocol simplifications. These modifications preserve the core security and authentication properties of the original design. In the following subsection, we formally state the underlying assumptions and provide justification for the soundness of these simplifications.

\subsection{Assumptions and Justification}

\subsubsection{Tag and Reader can be Combined Into One Entity}
All messages exchanged between the tag and the reader are also exchanged between the reader and the server. Therefore, if an adversary intends to eavesdrop on the communication between the tag and the reader, it can achieve the same by intercepting the reader-server communication. Hence, for the purpose of formal verification through AVISPA, the tag and the reader can be treated as a single combined entity. It verifies the logical consistency of the message sequence against classical DY threats, such as  MITM, replay, and impersonation, ensuring that the fundamental protocol flow is sound. In contrast, in our informal security analysis, the three part model where tag, reader and server are treated as separate entities across two insecure channels, addresses resilience to quantum attacks, which are beyond the scope of symbolic modeling tools. The informal analysis proves that the protocol maintains identity privacy, unlinkability, and unforgeability even when the reader is treated as a potentially untrusted intermediary.

\subsubsection{All Messages that are Related to ISIS problem can be Combined Into One Message Which is Always Safe Due to Hardness of ISIS}
We consider the set of ($\alpha, \beta, c_1, c_2, c_3, \gamma, \delta, c_6, c_7, c_8$) ten component messages as a single aggregated message. The message represents some arithmetic operations between session specific random nonces and long term private keys. These components are inherently interdependent and collectively necessary for successful authentication. If any component is altered by an adversary, the authentication process will fail. Therefore, for the purpose of analysis, we treat this collection of messages as a single, unified message. Any modification to any individual component is treated as a modification of the entire message, thereby resulting in authentication failure. Due to the hardness of the ISIS problem, an adversary is unable to derive the long-term secret keys of the reader and the tag from the component messages. AVISPA employs a symbolic model. While it confirms that the protocol logic is sound under the DY model, it relies on the well-known hardness of the ISIS problem to claim quantum resistance. 

\subsubsection{Symmetric Key Encryption as Substitute for ISIS}
All messages exchanged between the server and the combined entity of tag and reader are encrypted using a symmetric key encryption scheme. This symmetric key represents ISIS based encryption; this substitution is made as the actual matrix multiplications and arithmetic operations involved in ISIS are not directly supported within the AVISPA verification framework. This represents an approximation because the DY model assumes that these primitives are cryptographically perfect `black boxes'. While this abstraction does not capture potential vulnerabilities arising from the algebraic structure of the underlying lattice, it provides a rigorous proof that the protocol's message sequence and binding mechanisms are logically sound. The mathematical security against sub-symbolic or quantum attacks is separately established by the known hardness of the ISIS problem.

\subsection{Implementation of Protocol in AVISPA}
The HLPSL code in Listing \ref{lst:TR} defines a role \texttt{TR}. This role represents the combination of the tag and the reader as a single entity. The role interacts with another role, \texttt{S}, which represents the server.

This role progresses through multiple states, governed by a local variable \texttt{State}, which is initialized to 0. It also uses three message-related variables: \texttt{X2}, \texttt{M}, and \texttt{Xs}, all of type \texttt{text}.

The protocol starts with state 0, where upon receiving the message \texttt{start}, it generates fresh values \texttt{M'} and \texttt{X2'}, and sends an encrypted message containing these values using a shared symmetric key \texttt{K}. These values are declared as the secrets between the involved parties and labeled as \texttt{sec\_1} and \texttt{sec\_2}.

\begin{lstlisting}[caption={The listing shows our AVISPA implementation for the tag-reader role, where we consider the tag and the reader as a single entity.},label={lst:TR}]
role role_TR(TR:agent,S:agent,K:symmetric_key,SND,RCV:channel(dy))
played_by TR
def=
	local
		State:nat,X2:text,M:text, Xs : text
	init
		State := 0
	transition
		1. State=0 /\ RCV(start) =|> 
                   State':=1 /\  M' := new() /\ X2':=new()  /\ SND({M'.X2'}_K) 
			/\ secret(M',sec_1,{TR,S})
			/\ secret(X2',sec_2,{TR,S})

		2. State=1 /\ RCV({Xs'}_K) =|> 
                   	State':=2 /\ SND({Xs'.X2}_K) 
			/\ witness(TR,S,auth_1,Xs')
        
        
                3. State=2 /\ RCV({X2}_K) =|>
			State':=3
                    /\ request(TR,S,auth_2,X2)
			
end role
\end{lstlisting}

Next, the role advances to state 1, where it receives an encrypted message containing \texttt{Xs'}. In response, it sends back an encrypted pair, \texttt{Xs'} and \texttt{X2}. This step establishes a \texttt{witness} event, asserting that the authenticity of the sender has been verified through the value \texttt{Xs'}, labeled as \texttt{auth\_1}.

Finally, in state 2, the role receives an encrypted message containing \texttt{X2}, and triggers a \texttt{request} event indicating a request to authenticate the sender using the received value. This authentication request is labeled as \texttt{auth\_2}.

The HLPSL code in Listing \ref{lst:server} defines the role \texttt{S} for the server and uses the variable \texttt{State} to track its progress and involves temporary message variables \texttt{Xs}, \texttt{X2}, and \texttt{M}, all of type \texttt{text}. Initially in state 0, it receives an encrypted message \(\{M'.X2'\}_K\) from \texttt{TR}, generates a fresh nonce \texttt{Xs'}, and replies with \(\{Xs'\}_K\), encrypted using the shared key \texttt{K}. The value \texttt{Xs'} is declared secret between \texttt{TR} and \texttt{S}, and labeled \texttt{sec\_3}. In state 1, it receives \(\{Xs'.X2\}_K\), responds with \(\{X2\}_K\), and triggers a \texttt{request} and a \texttt{witness} event to assert mutual authentication, which are labeled \texttt{auth\_1} and \texttt{auth\_2}, respectively.

\begin{lstlisting}[caption={The listing shows our AVISPA implementation for the server role.},label={lst:server}]
role role_S(S:agent,TR:agent,K:symmetric_key,SND,RCV:channel(dy))
played_by S
def=
	local
		State:nat,Xs:text,X2:text,M:text
	init
		State := 0
	transition
		1. State=0 /\ RCV({M'.X2'}_K) =|> 
                   State':=1 /\ Xs' := new() /\ SND({Xs'}_K)
			/\ secret(Xs',sec_3,{TR,S})

		2. State=1 /\ RCV({Xs'.X2}_K) =|> 
                   State':=2 
			/\ request(S,TR,auth_1,Xs')
                    /\ SND({X2}_K) 
			/\ witness(S,TR,auth_2,X2)
end role
\end{lstlisting}

In Listing~\ref{lst:session}, we define a session which creates two channels, \texttt{SND} and \texttt{RCV} (dy = Dolev-Yao intruder model), and composes both the roles \texttt{TR} and \texttt{S}, allowing them to interact in a single session. The \texttt{environment} role sets up the initial configuration for AVISPA’s analysis.

\begin{lstlisting}[caption={The listing shows our AVISPA implementation for session setup, where both the tag-reader role and the server role are intialized.},label={lst:session}]
role session(TR:agent,S:agent,K:symmetric_key)
def=
	local
		SND,RCV:channel(dy)
	composition
		role_TR(TR,S,K,SND,RCV) /\ role_S(S,TR,K,SND,RCV)
end role
\end{lstlisting}

Listing~\ref{lst:env} defines constants such as agents \texttt{tr} and \texttt{s}, a symmetric key \texttt{k}, and claim identifiers. The intruder is assumed to know the agent names. The protocol is initiated with \texttt{session(tr, s, k)} and run twice to simulate a replay attack.

\begin{lstlisting}[caption={The listing shows our AVISPA implementation of the environment, where we declare agents, symmetric\_keys, and protocol\_id, and intialize a session.}, label={lst:env}]
role environment()
def=
	const
		k:symmetric_key,
                tr,s:agent,
                auth_1,auth_2,sec_1,sec_2, sec_3:protocol_id
	intruder_knowledge = {tr,s}
	composition
		session(tr,s,k) /\ session(tr,s,k)
end role
\end{lstlisting}

The \texttt{goal} in Listing~\ref{lst:goals} outlines the security properties that the protocol is designed to achieve. It ensures mutual authentication through the claims

\begin{lstlisting}[caption={The listing declares the authentication and secrecy goals of the protocol.},label={lst:goals}]
goal
    authentication_on auth_1,auth_2
    secrecy_of sec_1,sec_2,sec_3
end goal
\end{lstlisting}

 \texttt{auth\_1} and \texttt{auth\_2}, thereby confirming that both parties can verify each other’s identity. Additionally, it guarantees the secrecy of the values labeled \texttt{sec\_1}, \texttt{sec\_2}, and \texttt{sec\_3}, preserving the confidentiality of the exchanged information.

\subsection{Results}

\begin{figure}
    \centering
    \includegraphics[width=0.95\linewidth]{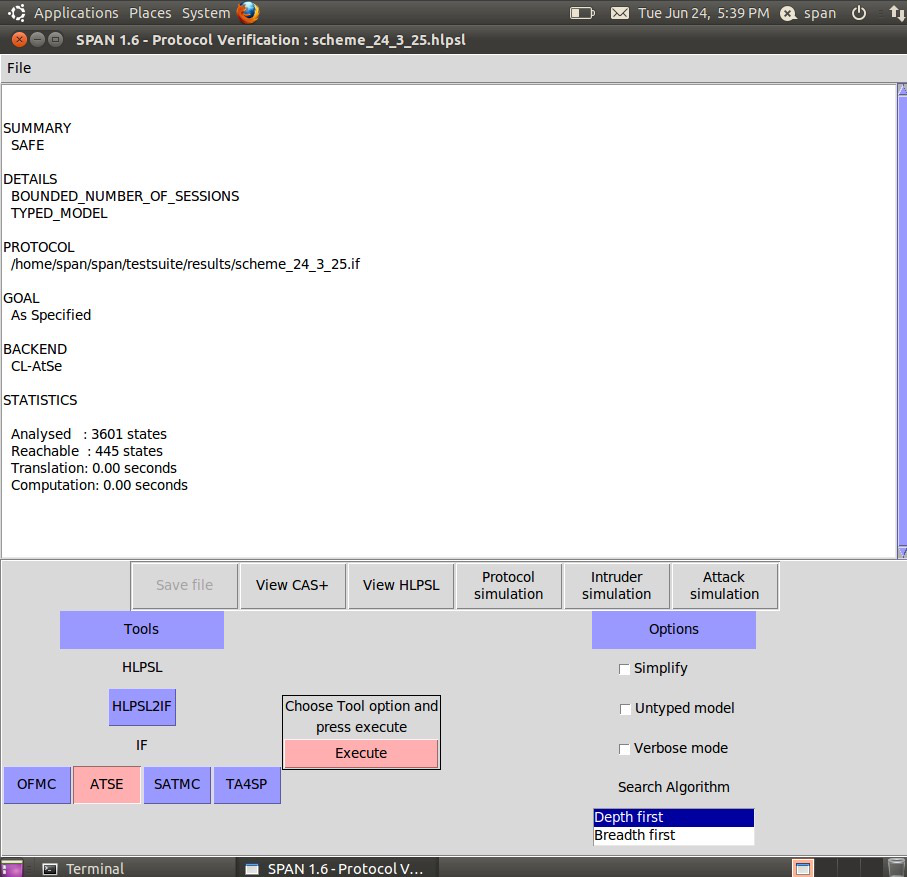}
    \caption{The figure shows the results obtained using the AVISPA ATSE tool.}
    \label{avispa-result}
\end{figure}

Fig.~\ref{avispa-result} shows the outcome of the verification process, which indicates that the protocol is SAFE, meaning that it is secure against the class of attacks modeled and tested by the AVISPA ATSE framework. This result confirms that the protocol resists common security threats such as replay attacks, man-in-the-middle attacks, and impersonation.

\section{Performance Evaluation}\label{performance_evaluation}

Our protocol includes several tunable parameters: $m$, $n$, $q$, and $l$. We adopt the parameter set proposed in  \cite{Cayrel20123363}, which is selected based on the complexity of known attacks that recover short lattice vectors. These parameters ensure the desired security level against such attacks.

In our scheme, the security parameter $m$ defines the length of the binary vectors used as private keys. For a 100-bit security level (i.e., the attacker needs to perform $2^{100}$ operations to break the scheme), we set $m = 2048$.

The other parameters are set as follows:
\begin{itemize}
    \item $n = 64$: number of rows in $A_t$ and $A_r$,
    \item $q = 257$: a prime modulus defining the finite field $\mathbb{Z}_q$,
    \item $l = 256$: the bit-length of hashed outputs.
\end{itemize}

In practical implementations, we use a small prime modulus $q=257$ which fits well with $8$-bit arithmetic and is therefore suitable for resource-constrained RFID microcontrollers. We evaluate the performance of our protocol in terms of the storage, communication, and computation costs at all the participating entities versus $m$.

\subsection{Storage Cost}

The permutation matrices such as $P_t$ and $P_r$ are of size $m \times m$, but they do not need to be stored as full binary matrices. Since each permutation over $m$ elements can be represented by an ordered list of $m$ indices, and each index requires $\log_2 m$ bits, storing a permutation matrix requires only $m \log_2 m$ bits. Similarly, any element from the finite field $\mathbb{Z}_q$ can be represented using $\log_2 q$ bits, since there are $q$ possible values.

Table~\ref{tab:combined_storage_cost} summarizes the storage costs at the Tag, Reader, and Server. The total storage cost is
\[
4m \log(q) + 7m\log(m) + 2n\log(q) + 6nm\log(q).
\]

\begin{figure}
    \centering
    \includegraphics[width=0.95\linewidth]{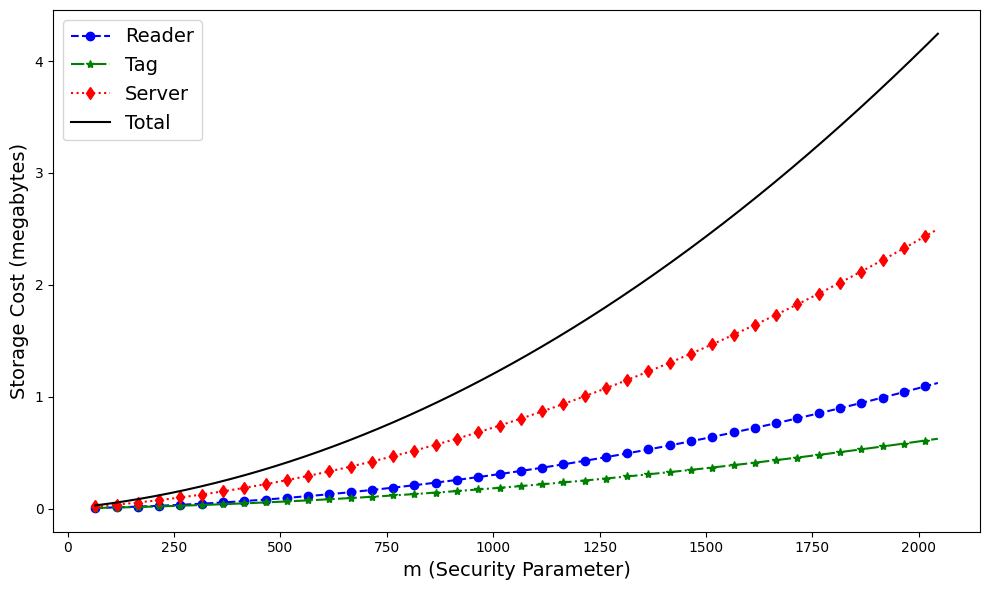}
\caption{The figure shows the variation of the storage costs at different entities versus \( m \). It also depicts the total storage cost of the proposed protocol.}
    \label{fig:storage}
\end{figure}

Fig.~\ref{fig:storage} illustrates the storage requirements of various entities under the proposed scheme versus the security parameter $m$. For  $m = 2048$, the storage cost is 0.129 MB for the tag, 0.132 MB for the reader, and  0.515 MB for the server.

For practical implementation, the permutation matrices $P_t$ and $P_r$ are not stored as full $2048 \times 2048$ bit matrices (which would take $512$ KB). Instead, they are stored as ordered index lists of $m$ elements. Since $\log_2(2048) = 11$, each index takes $11$ bits, resulting in a total storage requirement of only $\approx 2.8$ KB. Moreover, the tag does not need to store $A_t$ statically. Instead, it can store a small $256$-bit seed and regenerate $A_t$ dynamically using a lightweight extendable-output function (XOF), effectively reducing the storage of $A_t$ from $147$ KB to $32$ bytes. As $x_t$ requires roughly $2.3$ KB, the total non-volatile memory required on the tag is reduced to approximately  $5$ KB. So, the protocol is feasible on the target RFID platform.

\begin{table}[htbp]
\caption{The table presents the storage costs incurred at the tag, reader, and server under the proposed scheme.}
\renewcommand{\arraystretch}{1.1}
\centering
\begin{tabular}{|l|l|l|}
\hline
\textbf{Entity} & \textbf{Parameter} & \textbf{Cost (bits)} \\
\hline
\multirow{3}{*}{Tag} 
& $ x_t $   & $ m\log(q) $ \\
& $ P_t $   & $ m\log(m) $ \\
& $ A_t $   & $ nm\log(q) $ \\
\cline{2-3}
& \textbf{Total} & $ m\log(q) + m\log(m)$ \\ 
& & $+ nm\log(q) $ \\
\hline

\multirow{3}{*}{Reader} 
& $ x_r $   & $ m\log(q) $ \\
& $ P_r,P_r^{-1} $   & $ m\log(m) $ \\
& $ A_r $   & $ nm\log(q) $ \\
\cline{2-3}
& \textbf{Total} & $ m\log(q) + 2m\log(m)$ \\ 
& & $+ nm\log(q) $ \\
\hline

\multirow{6}{*}{Server} 
& $ x_r, x_t $   & $ m\log(q) $ \\
& $ y_r, y_t $   & $ n\log(q) $ \\
& $ P_r, P_t $   & $ m\log(m) $ \\
& $ P_r^{-1}, P_t^{-1} $   & $ m\log(m) $ \\
& $ A_r , A_t $   & $ nm\log(q) $ \\
& $ A_rP_r^{-1}, A_tP_t^{-1}  $   & $ nm\log(q) $ \\
\cline{2-3}
& \textbf{Total} & $ 2m\log(q) + 2n\log(q)$ \\
& & + $4m\log(m)+ 4nm\log(q) $ \\
\hline
\end{tabular}

\label{tab:combined_storage_cost}
\end{table}

\begin{figure}
    \centering
    \includegraphics[width=0.95\linewidth]{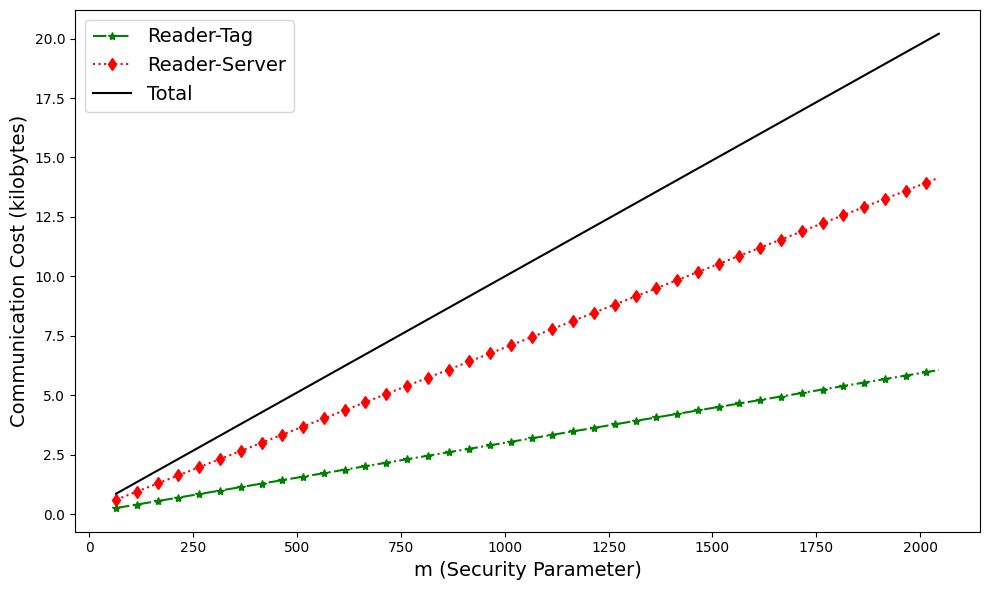}
\caption{The figure shows the variation of the communication costs of different channels versus \( m \). It also depicts the total communication cost of the proposed protocol.}
    \label{fig:communication}
\end{figure}

\subsection{Communication Cost}

Table \ref{tab:combined_comm_cost} summarizes the communication costs between different entities. The total communication cost is 
\[
= 10m\log(q)+3\log(q) + 7l.
\]

Fig. \ref{fig:communication} illustrates the communication requirements of various channels under the proposed scheme versus the security parameter $m$. For  $m = 2048$, the communication cost for the reader-tag channel is approximately 6.06 KB and for the reader-server channel is around 14.16 KB.

\begin{table}[htbp]
\caption{The table shows the communication costs among the tag, reader, and server under the proposed protocol.}
\renewcommand{\arraystretch}{1.1}
\centering
\begin{tabular}{|p{2.4cm}|p{2.4cm}|p{2.6cm}|}
\hline
\textbf{Channel} & \textbf{Parameter} & \textbf{Cost (bits)} \\
\hline

\multirow{2}{=}{Reader to Tag (1)} 
& $ \alpha $ & $ \log(q) $ \\
\cline{2-3}
& \textbf{Total} & $ \log(q) $ \\
\hline

\multirow{3}{=}{Tag to Reader} 
& $ c_1, \beta, c_3 $ & $ m\log(q) $ \\
& $ c_2 $ & $ l $ \\
\cline{2-3}
& \textbf{Total} & $ 3m\log(q) + l $ \\
\hline

\multirow{4}{=}{Reader to Server (1)} 
& $ \alpha, \gamma $ & $ \log(q) $ \\
& $ \beta, \delta, c_1, c_3, c_6, c_8 $ & $ m\log(q) $ \\
& $ c_2, c_7 $ & $ l $ \\
\cline{2-3}
& \textbf{Total} & $ 6m\log(q) + 2l + 2\log(q) $ \\
\hline

\multirow{2}{=}{Server to Reader (1)} 
& $ c_{11} $ & $ m\log(q) $ \\
\cline{2-3}
& \textbf{Total} & $ m\log(q) $ \\
\hline

\multirow{2}{=}{Reader to Server (2)} 
& $ c_{13} $ & $ l $ \\
\cline{2-3}
& \textbf{Total} & $ l $ \\
\hline

\multirow{2}{=}{Server to Reader (2)} 
& $ c_{14},c_{15} $ & $ l $ \\
\cline{2-3}
& \textbf{Total} & $ 2l $ \\
\hline

\multirow{2}{=}{Reader to Tag (2)} 
& $ c_{14} $ & $ l $ \\
\cline{2-3}
& \textbf{Total} & $ l $ \\
\hline

\end{tabular}

\label{tab:combined_comm_cost}
\end{table}

\subsection{Computation Cost}

Generating a random vector is assumed to take a number of operations that is linear in its size. Generating a random number from \( \mathbb{Z}_q \) is assumed to cost 1 operation. Fixed-length hash functions such as \( H_1 \) (e.g., SHA-256) are assumed to incur a cost linear in the input size, while hash functions such as \( H_2 \), which produce variable-length outputs (of size \( m - n \)), are taken to have cost equal to the sum of the input and output sizes. Multiplying with a permutation matrix is counted as $m$ operations, as it corresponds to a simple reordering of the input vector. 

\begin{table}[htbp]
\caption{The table presents the computation costs incurred at the tag, reader, and server under the proposed scheme.}
\renewcommand{\arraystretch}{1.1}
\centering
\begin{tabular}{|p{1.6cm}|p{2.8cm}|p{3.0cm}|}
\hline
\textbf{Entity} & \textbf{Computation} & \textbf{Cost (operations)} \\
\hline

\multirow{5}{*}{Tag} 
& $ x_1, u_t $ & $ m $ \\
& $ \beta $ & $ 3m $ \\
& $ c_1,c_3 $ & $ 2m $ \\
& $ c_2 $ & $ 2mn + n + m $ \\
& $c_{14}$ & $3m$ \\
\cline{2-3}
& \textbf{Total} & $ 13m + n + 2mn $ \\
\hline

\multirow{6}{*}{Reader} 
& $ \alpha, \gamma $ & $ 1 $ \\
& $ x_2, u_r $ & $ m $ \\
& $ c_6,c_7,c_8, c_{12},c_{13} $ & $ 2m $ \\
& $ \delta, c_{15} $ & $ 3m $ \\
\cline{2-3}
& \textbf{Total} & $ 18m + 2 $ \\
\hline

\multirow{6}{*}{Server} 
& $ c_9,c_4$ & $  2n + 2mn$ \\
& $ c_{10}, c_5, c_2, c_7, c_{13} $ & $ 2m $ \\
& $ x_s$ & $ m $ \\
& $ c_{11},c_{14},c_{15} $ & $ 4m$ \\
\cline{2-3}
& \textbf{Total} & $ 23m + 4n + 4mn $ \\
\hline

\end{tabular}

\label{tab:computation_cost}
\end{table}

\begin{figure}
    \centering
    \includegraphics[width=1\linewidth]{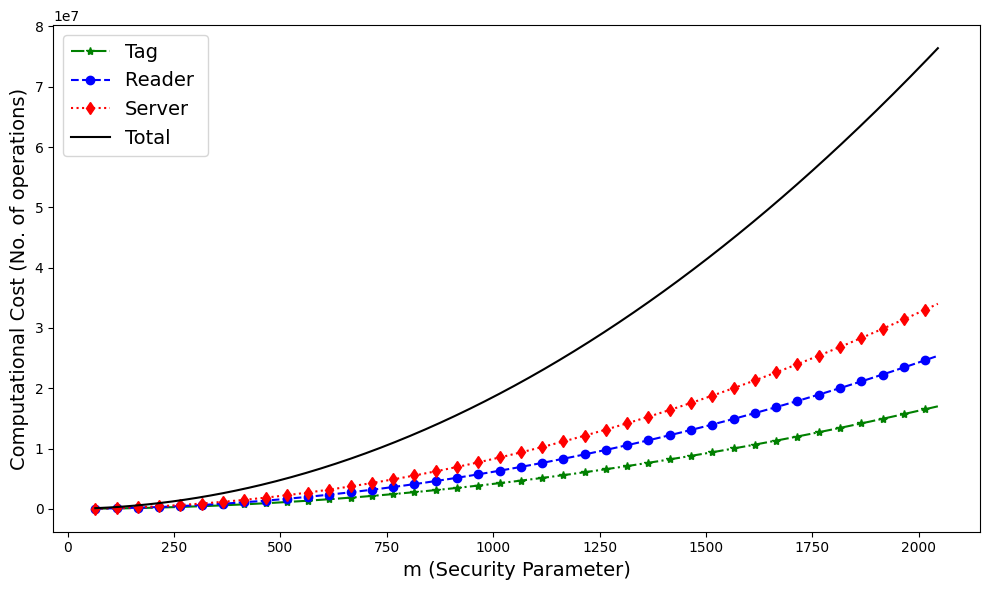}
    \caption{The figure shows the variation of the computation cost at different entities versus \( m \). It also depicts the total computation cost of the proposed protocol.}
    \label{fig:computation}
\end{figure}

Table~\ref{tab:computation_cost} summarizes the computation costs at the Tag, Reader, and Server in terms of the number of operations. The total computation cost is
\[
54m +5n + 6mn +2.
\]
Fig. \ref{fig:computation} illustrates the computation requirements of different entities under the proposed scheme versus the security parameter $m$.

During the practical implementation of our scheme, the protocol requires the tag to generate an $m=2048$ dimensional random vector during each authentication session, but the tag does not need to generate these random bits from its physical entropy source. Instead, it can generate a small $256$-bit seed and expand it into the $m=2048$ vector using a lightweight Cryptographically Secure Pseudo-Random Number Generator (CSPRNG). Also, some of this randomness can be pre-calculated during the tag's ``idle" time between scans to reduce the computation time.

\section{Comparison of Security Properties with Existing Schemes} \label{comparison}

Table \ref{tab:security_properties} shows a comparison of the proposed scheme with various schemes proposed in prior work in terms of security properties. This comparison is in terms of a number of security properties, such as location privacy, unforgeability, scalability,  MITM, replay, impersonation, and quantum attacks, and  support for insecure reader-server channels. The schemes proposed in \cite{CHO201558, DASS2016100, dong2018cloud, inproceedings, RePEc:sae:intdis:v:15:y:2019:i:7:p:1550147719862223,provable} are not secure against quantum attacks. All the schemes from prior work \cite{ghosh, CHO201558, DASS2016100, dong2018cloud, inproceedings, RePEc:sae:intdis:v:15:y:2019:i:7:p:1550147719862223, provable,base} included in the table  assume that the reader-server channel is secure. In contrast, our proposed scheme is secure even when the reader-server channel is insecure; it also satisfies all the other security requirements needed for RFID systems.

\begin{table*}
\centering
\caption{The table provides a comparison of the security properties of the proposed scheme and those of various existing protocols. A checkmark ($\checkmark$) indicates that the corresponding scheme satisfies the security property, whereas a cross ($\times$) indicates that it does not.}
\label{tab:security_properties}

\setlength{\tabcolsep}{3pt}
\renewcommand{\arraystretch}{1.1}

\begin{tabularx}{\textwidth}{>{\raggedright\arraybackslash}X *{10}{c}}
\hline
\textbf{Security Properties} & \textbf{\cite{Chien2010}} & \textbf{\cite{inproceedings}} & \textbf{\cite{CHO201558}} & \textbf{\cite{DASS2016100}} & \textbf{\cite{dong2018cloud}} & 
\textbf{\cite{RePEc:sae:intdis:v:15:y:2019:i:7:p:1550147719862223}} &
\textbf{\cite{provable}} &
\textbf{\cite{ghosh}} &
\textbf{\cite{base}} & 
\textbf{Ours}\\ 
\hline
Identity/ Location Privacy    & $\times$ & $\times$ & $\times$ & $\checkmark$ & $\checkmark$ & $\checkmark$ & $\checkmark$ & $\checkmark$&$\checkmark$ & $\checkmark$\\ \hline
Unforgeability      & $\checkmark$ & $\checkmark$ & $\checkmark$ & $\checkmark$ & $\times$ & $\checkmark$ & $\checkmark$ & $\checkmark$& $\checkmark$  & $\checkmark$\\ \hline
Scalability         & $\times$ & $\times$ & $\times$ & $\times$ & $\times$ & $\checkmark$ & $\checkmark$ & $\checkmark$& $\checkmark$ & $\checkmark$\\ \hline
MITM Attack         & $\checkmark$ & $\checkmark$ & $\checkmark$ & $\checkmark$  & $\checkmark$ & $\checkmark$ & $\checkmark$ & $\checkmark$& $\checkmark$  & $\checkmark$\\ \hline
Replay Attack       & $\checkmark$ & $\checkmark$ & $\checkmark$ & $\checkmark$  & $\checkmark$ & $\checkmark$ & $\checkmark$ & $\checkmark$& $\checkmark$  & $\checkmark$\\ \hline
Impersonation Attack& $\checkmark$ & $\checkmark$ & $\checkmark$ & $\checkmark$  & $\checkmark$ & $\checkmark$ & $\checkmark$ & $\checkmark$& $\checkmark$  & $\checkmark$\\ \hline
Quantum-Resistant   & $\times$ & $\times$ & $\times$ & $\times$ & $\times$ & $\times$ & $\times$ & $\checkmark$& $\checkmark$ & $\checkmark$ \\ \hline
Insecure Reader-Server Channel Supported & $\times$ & $\times$ & $\times$ & $\times$ & $\times$ & $\times$ & $\times$ & $\times$& $\times$ & $\checkmark$ \\ \hline
\end{tabularx}
\end{table*}

\section{Conclusions and Future Work} \label{conclusion}
We proposed a novel quantum-resistant authentication scheme, which uses lattice-based cryptography and leverages the hardness of the ISIS problem, for RFID systems.  We demonstrated the correctness of our proposed scheme and showed that it is secure even when the tag-reader and reader-server channels are both insecure.  We presented a detailed semi-formal security analysis, which shows that our scheme provides robust security against MITM, impersonation, reflection, and replay attacks, while also ensuring unforgeability and preserving anonymity. We also provided a formal security analysis using the AVISPA tool. We evaluated the performance of the proposed scheme in terms of its storage, communication, and computation costs and compared its security properties with those of several schemes proposed in prior work.  To the best of our knowledge, this paper is the first quantum-resistant authentication protocol for RFID systems that comprehensively addresses the insecurity of both the reader-server and tag-reader communication channels.

Some directions for future research include optimizing the storage, communication, and computational efficiency of the proposed scheme, and implementing it on  lightweight hardware platforms.

\bibliographystyle{elsarticle-num}
\bibliography{cas-refs}


\end{document}